\documentclass[%
 reprint,
showpacs,preprintnumbers,
 amsmath,amssymb,
 aps,
]{revtex4-1}

\usepackage{graphicx}
\usepackage{dcolumn}
\usepackage{bm}
\usepackage{epstopdf}

\begin{document}


\title{Indirect RKKY interaction in armchair graphene nanoribbons}

\author{Karol Sza\l{}owski}

\email{kszalowski@uni.lodz.pl}


 \affiliation{Department of Solid State Physics, University of \L{}\'od\'z, ulica Pomorska 149/153, PL90-236 \L{}\'od\'z, Poland}

\date{\today}

\begin{abstract}
A form of an indirect Ruderman-Kittel-Kasuya-Yosida (RKKY)-like coupling between magnetic on-site impurities in armchair graphene nanoribbons is studied theoretically. The calculations are based on a tight-binding model for a finite nanoribbon system with periodic boundary conditions. A pronounced Friedel-oscillation-like dependence of the coupling magnitude on the impurity position within the nanoribbon resulting from quantum size effects is found and investigated. In particular, the distance dependence of coupling is analysed. For semiconducting nanoribbons, this dependence is exponential-like, resembling the Bloembergen-Rowland interaction. In particular, for metallic nanoribbons, interesting behaviour is found for finite length systems, in which zero-energy states make an important contribution to the interaction. In such situation, the coupling decay with the distance can be then substantially slower. 

\end{abstract}

\pacs{75.30.Hx, 75.75.-c, 75.30.Et, 73.22.Pr}
\maketitle


\section{\label{sec:level1}Introduction}

Graphene, being an unique two-dimensional material \cite{Geim2004,Novoselov2005}, opens new possibilities for contemporary physics and technology. One of the challenges is finding route to the spintronic applications of graphene and its derivatives\cite{Wees2011,Yazyev2010}, what motivates experimental and theoretical studies of their magnetic properties. In particular, interactions between magnetic moments introduced to graphene lattice focus considerable interest. One of the known mechanisms of interaction between magnetic moments embedded in the host material is the Ruderman-Kittel-Kasuya-Yosida (RKKY) mechanism \cite{Rudermann1954,Kasuya1956,Yosida1957} of indirect coupling mediated by the charge carriers. The form of RKKY interaction, originally derived for bulk, metallic, three-dimensional systems, is sensitive to the geometry and dimensionality of the underlying system and closely connected with the electronic structure of the host material. This opened room for studies of RKKY coupling in low-dimensional systems\cite{Kittel1968,Litvinov1998,Aristov1997,Beal-Monod1987,Yafet1987,Giuliani2005,Szalowski2008,Bak2000}, in particular to mention thin films\cite{Wojtczak1969,Balcerzak2007,Balcerzak2007b} at the crossover between the three- and two-dimensions, or for pseudo-one-dimensional systems \cite{Mi2011}. In particular, in metallic, two-dimensional systems, RKKY coupling has been predicted to decay with the distance according to $J^{RKKY}\propto 1/r^2$. However, the peculiar dispersion relation for charge carriers in monolayer graphene\cite{CastroNeto2009}, being a zero-gap semiconductor, together with a bipartite nature of the underlying graphene crystalline lattice with dominant nearest-neighbours electron hopping, causes the indirect RKKY coupling in graphene to differ from that found in 2D metals. The indirect RKKY exchange in graphene attracts considerable attention \cite{Power2012,Lee2012,Lee2012b,Kogan2011,Kogan2012,Uchoa2011,Ferreira2011,Sherafati2011,Power2012b,Peng2012,Bunder2012,Annika2010a,Annika2010b,Dugaev2006,Brey2007,Wunsch2006,Saaremi2007,Jiang2012,Bunder2009,Sherafati2012,Szalowski2011,Cheianov2007,Braun2012}. It has been verified, that its asymptotic distance dependence is $J^{RKKY}\propto 1/r^3$ for monolayer graphene (somehow like in a three-dimensional system). Moreover, in undoped graphene, the coupling between magnetic impurities in the same sublattice is always ferromagnetic, while for the impurities in different sublattices, it is antiferromagnetic.

One of the promising building blocks for graphene spintronics are graphene nanoribbons. These nanostructures offer highly tuneable electronic structure, vitally dependent on its geometry, mainly the edge form. In general, the form of the edge of graphene and similar structures is known from early experimental and theoretical studies to shape electronic states in the vicinity of the Fermi level(e.g. \cite{Wakabayashi1996,Klusek2005}). Besides zigzag-edged nanoribbons, offering expected spontaneous edge magnetic polarization and thus concentrating the interest, also armchair-edged graphene nanoribbons (AGNRs) attract much attention (e.g.\cite{Power2011,Lee2012c,Lipinski2012}). 
The interest in properties of AGNR is especially motivated by, for example, recent achievements in fabrication of narrow nanoribbons characterized by well-controlled width\cite{Blankenburg2012,Ruffieux2012}. On the other hand, the form of RKKY interaction in metallic nanoribbons with quadratic dispersion relation for the charge carriers has been also a subject of a current study\cite{Mi2011} which emphasized the possibility of control over the coupling due to its position dependence. The last feature, resulting from the pronounced quantum size effects, combined with an uniqueness of graphene, can be expected to promote interesting behaviour of RKKY interaction in graphene nanoribbons. Moreover, in general the effects of dimensional cross-over from quasi-one-dimensional nanoribbons to two-dimensional infinite monolayer appear interesting. This serves as motivation for computational studies of indirect coupling between magnetic impurities in armchair graphene nanoribbons presented in the paper. 

\section{Theory}

\begin{figure}
\includegraphics[scale=0.75]{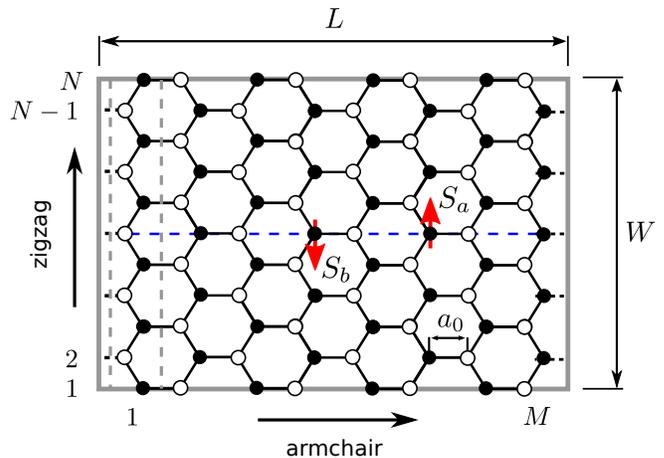}
\caption{\label{fig:fig1}Schematic view of the nanoribbon system used in the calculations. The solid box encloses the unit cell of width $W$ and length $L$, with $N$ carbon atoms at the zigzag edge. The dashed lines enclose a single row of atoms in zigzag direction; the unit cell is composed of $M$ such rows. Periodic boundary conditions connect the zigzag edges, so that $M+1\equiv 1$. The symmetry axis of the nanoribbon is marked with a dashed line. Open and filled circles denote the sites from two sublattices.}
\end{figure}

The basis of the calculations is the choice of a finite graphene system. The study was performed for a finite system, schematically depicted in Fig.~\ref{fig:fig1}. The unit cell used in calculations contained $N$ carbon atoms at the zigzag edge and the number of such zigzag atomic rows in armchair direction was equal to $M$, so in total the structure contained $M\cdot N$ atoms. Empty and filled circles allow to distinguish between the sites belonging to two sublattices. The row of atoms extending along armchair direction is composed of subsequent nearest-neighbour pairs (with the nearest-neighbour distence equal to $a_0$) and can be called a dimer line \cite{Alfonsi2012}); the system contains thus $N$ such dimer lines. The corresponding length of the nanostructure was $L=3a_{0}M/2$, while its width was equal to $W=\sqrt{3}a_{0}\left(N-1\right)/2$. Periodic boundary conditions were used to connect the zigzag edges, so that for the zigzag atomic rows $M+1\equiv 1$.  The usage of such a geometry had two aims, first one being to simulate the behaviour of the infinite length AGNR, the second one being to investigate some important finite-size effects discussed further. Only the nanoribbons for odd values of $N$ were considered, since the interest was focused on the AGNRs possessing the symmetry axis. The symmetry axis of the AGNR was described by $\delta=0$, while the armchair edges are shifted by $\delta=\pm\left(N-1\right)/2$ dimer lines from the center. 

The values of $M$ used in the calculations of distance dependences of the coupling (Figs.~\ref{fig:fig3},\ref{fig:fig5}) were $M=402$, while the rest of results, apart from those illustrated in Fig.~\ref{fig:fig4}(b) was obtained using $M=202$. It has been verified that in absence of zero-energy states participating in indirect coupling (see further discussion), such a values of $M$ were large enough to guarantee the convergence of the results to the limit of infinite-length nanoribbon. Anticipating the further analysis, let us mention that the largest distance between the magnetic impurities considered in this paper was $r/a_0=27$, so that the padding around the impurity pair in the armchair direction was at least more than 10 times larger than that distance (see the results and discussion by Black-Schaffer\cite{Annika2010a}, demonstrating that the padding as twice as the distance between the impurities is sufficient). 

The electronic structure of finite AGNRs was modeled using a tight-binding Hamiltonian for $p^{z}$ electrons, with nearest-neighbours hopping $t$, supplemented with an Anderson-Kondo impurity term \cite{Annika2010a}:
\begin{align}
\label{eq:eq1}
\mathcal{H}&=-t\sum_{\left\langle i,j\right\rangle ,\sigma}^{}{\left(c^{\dagger}_{i,\sigma}c_{j,\sigma}+c^{\dagger}_{j,\sigma}c_{i,\sigma}\right)}\nonumber\\&+\frac{J}{2}S^{z}_{a}\left(c^{\dagger}_{a,\uparrow}c_{a,\uparrow}-c^{\dagger}_{a,\downarrow}c_{a,\downarrow}\right)+\frac{J}{2}S^{z}_{b}\left(c^{\dagger}_{b,\uparrow}c_{b,\uparrow}-c^{\dagger}_{b,\downarrow}c_{b,\downarrow}\right),
\end{align}
where $\left\langle i,j\right\rangle$ denotes summation over nearest neighbours in the lattice, while $\sigma=\,\uparrow,\downarrow$ is the electron spin.

\begin{figure}
\includegraphics[scale=0.56]{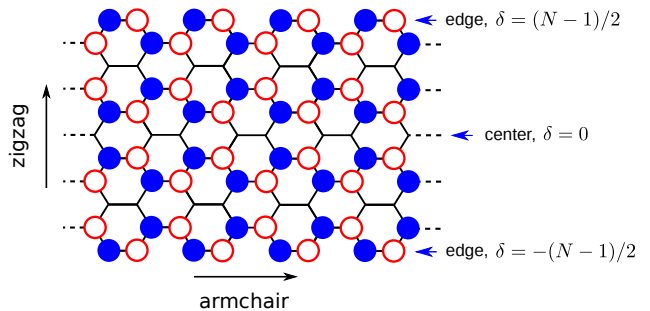}
\caption{\label{fig:fig2}Plot of the charge density associated with zero-energy states which are present in the electronic structure of metallic AGNR. Empty and filled circles depict the charge density for two distinct zero-energy states associated with two sublattices. The symmetry axis of the nanoribbon corresponds to $\delta=0$, while at the armchair edges $\delta=\pm \left(N-1\right)/2$. }
\end{figure}

The coupling between on-site impurity spins $S_a$ and $S_b$ (having the magnitude of $S=1$ for simplicity and located at sites $a$, $b$) and the electron spins at the same sites is parametrized by the energy $J$, called further a contact potential. In all the further results, except those presented in Fig.\ref{fig:fig4}(a), the value of $J/t=0.1$ was accepted. We note that the coupling is insensitive to the sign of $J$. Let us mention that for the sake of simplicity of numerical calculations, the Ising form of coupling between spins was assumed, and this assumption does not influence the resulting value of indirect coupling integral between impurity spins \cite{Szalowski2011}. The reduced Hamiltonian for impurity spin pair would be $H_{ab}=-J^{RKKY}\,S^z_{a}\,S^z_{b}$. In our convention the negative coupling values correspond to antiferromagnetic coupling, while the positive ones to ferromagnetic coupling. 

The Hamiltonian was diagonalized in single-particle approximation, with the help of LAPACK library \cite{Lapack}, yielding a set of electronic eigenenergies $\epsilon_{i}$ sorted in ascending order. The ground state energy (at $T=0$) for the system depended on the orientation of the impurity spins and was expressed by a sum of $M\cdot N$ least eigenvalues, $E\left(S_{a},S_{b}\right)=\sum_{i=1}^{M\cdot N}{\epsilon_{i}}$. This followed from the half-filling (charge neutrality) condition (one $p^{z}$ electron per each carbon atom). Then, the indirect coupling constant $J^{RKKY}$ was calculated from the expression\cite{Annika2010a,Szalowski2011}:
\begin{equation}
\label{eq:eq2}
2S^2J^{RKKY}=E\left(S_{a}=1,S_{b}=-1\right)-E\left(S_{a}=1,S_{b}=1\right).
\end{equation}

Note that the approach used allows for non-perturbative determination of indirect exchange integrals. 

Let us clarify here that we refer to the resulting indirect coupling as RKKY interaction. In general, as shown further, the coupling may contain a highly significant contribution resulting from the mechanism which can be explained by means of first order perturbation calculus, in contrast to the original RKKY mechanism involving a second-order process. However, for brevity we call the total resulting coupling the RKKY coupling\cite{Szalowski2011}. 

We mention that due to lack of free zigzag edges in the assumed geometry of AGNRs, no additional zero energy states localized at zigzag edges are introduced this way to the electronic spectrum.

The electronic structure of the AGNRs within nearest-neighbour tight-binding approximation has been extensively studied in the works of Wakabayashi \emph{et al.} \cite{Wakabayashi1999,Wakabayashi2009,Wakabayashi2010}. One of the results was that the AGNRs can be either metallic (M AGNRs), for widths $N=3n-1$ ($n=1,2,\dots$), or semiconducting (SC AGNRs), for $N=3n$ and $N=3n+1$. The SC AGNRs have an energy gap inversely proportional to the width. This behaviour was fully confirmed in our calculations. Let us note that the two series of SC AGNRs, $3n$ and $3n+1$, are inquivalent, so that two SC AGNRs from different series can display some differences in the properties other than resulting merely from the different gap value. Such effects were for example predicted for excitonic spectra \cite{Alfonsi2012,Jia2011,Mohammadzadeh2011}. As $N\to \infty$, the properties of AGNRs tend to that of an infinite graphene monolayer, being a zero-gap semiconductor. Let us emphasize that in our study we focus our interest only on the AGNRs with symmetry axis, therefore, the allowed values of $N$ are only odd. 

\subsection{Zero-energy states}      

Let us focus on metallic AGNRs. In absence of the magnetic impurities, the electronic structure of metallic ANGRs is characterized by the existence of two zero-energy states (each one with an additional two-fold spin degeneracy). The distribution of the charge density associated with both states is depicted in Fig.~\ref{fig:fig2} for an AGNR with $N=11$. First let us note that each of the states involves one sublattice only. Then, it follows that the wavefunction amplitudes of both zero-energy states vanish for specific dimer lines, namely for that characterized by the distance from AGNR center equal to $\delta=3d$, $d=0,1,\dots$. The number of such sites in our finite system is $MN/3$. In particular, the wavefunctions always vanish along the AGNR symmetry axis. The partial charges at the remaining $MN/3$ lattice sites of each sublattice are distributed uniformly and the probability of finding an electron is then equal to $n_0=3/\left(MN\right)$ for each of this sites, as follows from the normalization condition. Especially, the wavefunction amplitudes take nonzero values at the AGNR edge and for the neighbouring dimer line. The form of the electron wavefunctions for AGNRs has been also discussed recently by Sasaki \emph{et al.}\cite{Wakabayashi2011} (see the charge density for Dirac singularity zero-energy states sketched in the Fig.~1(b) in the Ref.~\onlinecite{Wakabayashi2011}, however, for a nanoribbon with even $N$). The wavefunction behaviour is a result of the interference of the electron wavefunctions incident and reflected from the armchair edge. The presence of similar interference effects has been also confirmed experimentally by means of STM/STS studies of the graphene sheets in the vicinity of the armchair edge\cite{Yang2010,Tian2011}.

\section{Results}

\begin{figure*}
\includegraphics[scale=0.25]{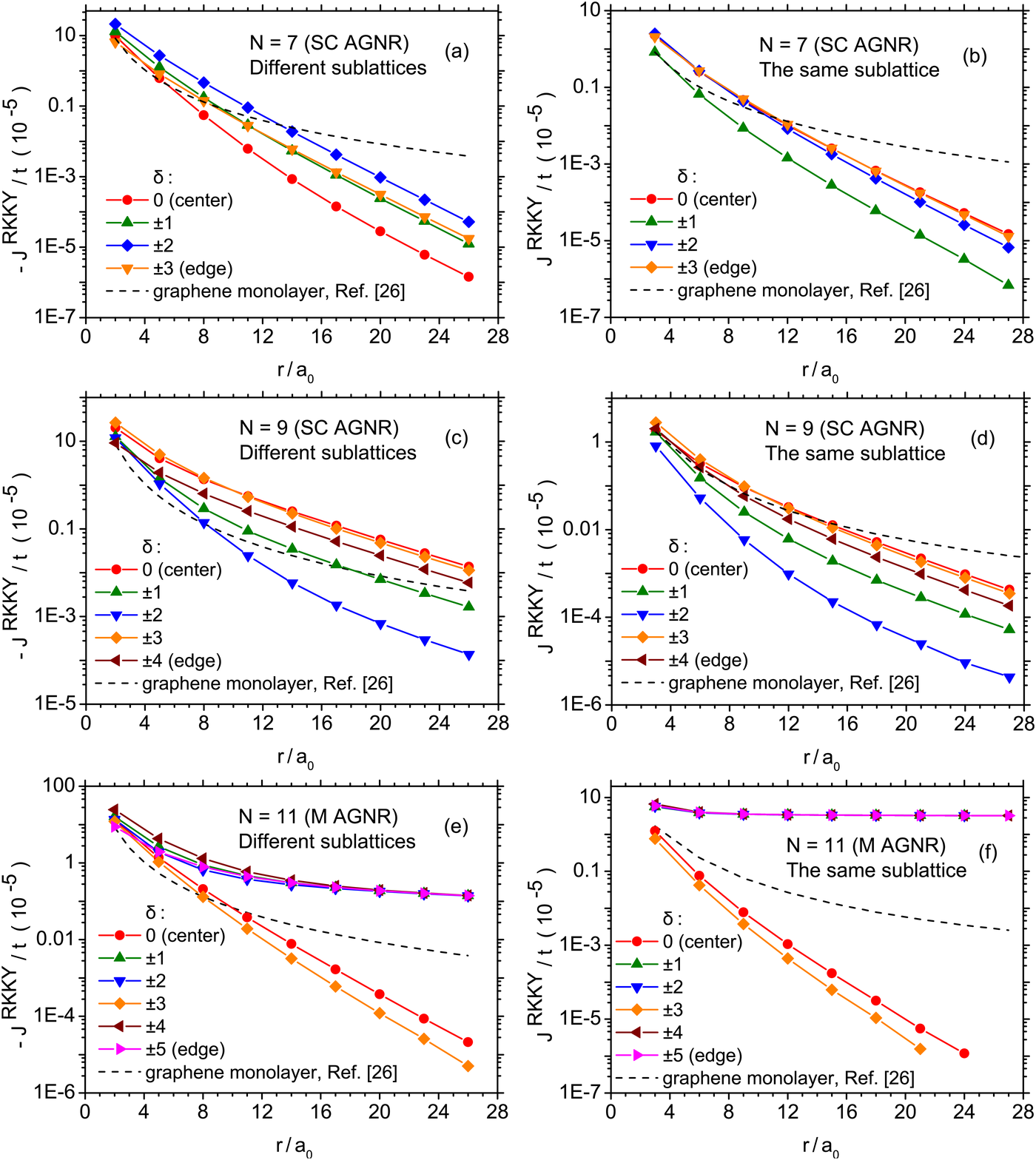}
\caption{\label{fig:fig3}The distance dependence of an indirect RKKY coupling between magnetic impurities situated in different sublattices (left) and the same sublattice (right), for two classes of SC AGNRs [(a)-(d)] and M AGNR [(e), (f)]. Dashed lines denote the results of perturbative calculation after the Ref.~\onlinecite{Sherafati2011}. The plots are presented for various distances of the impurities from the AGNR center, covering the whole range between the center and the edge of the nanoribbon.}
\end{figure*}

In this section we present the results of numerical calculations performed for our system. 
 
One of the consequences of the peculiar Dirac-like dispersion relation for charge carriers in graphene monolayer is that the asymptotic distance dependence of the RKKY coupling between magnetic impurities is $1/r^3$ (see e.g. \cite{Kogan2011,Sherafati2011,Annika2010a}), contrary to $1/r^2$ characteristic of two-dimensional metallic systems\cite{Beal-Monod1987,Litvinov1998,Aristov1997}. On the other hand, for metallic one-dimensional metallic systems the asymptotic dependence on $1/r$ is expected\cite{Kittel1968,Litvinov1998,Giuliani2005,Yafet1987}. Let us then analyze the distance dependence trends for the indirect coupling between magnetic impurities in AGNRs. The distance dependences are presented, for representative cases of three AGNRs with the widths of $N=7$, $9$ and $11$, in the Fig.~\ref{fig:fig3}. The separate plots are presented for two semiconducting AGNRs and for a metallic AGNR; on the other hand the coupling is also plotted separately for the impurities in the same sublattice and in different sublattces. Let us note that in graphene, due to the bipartite nature of its lattice, the coupling between impurities in the same sublattice is always ferromagnetic, while for the impurities in different sublattices it is antiferromagnetic.\cite{Saaremi2007} This rule is also obeyed in the case of AGNRs and no terms are present in the Hamiltonian to cause its breaking. The coupling energies are depicted in semilogarithmic scale to illustrate better the dependence ranging over many orders of magnitude, thus the plots present the absolute values of the couplings. In all the plots, the dashed lines present the results of the perturbative calculation of the RKKY coupling energy for armchair direction in monolayer graphene for $J/t=0.1$ performed using the Eq.~(19) and Eq.~(26) in the Ref.~\onlinecite{Sherafati2011} (for the same and for different sublattices, respectively).  

First let us observe that for semiconducting AGNR with $N=7$, the coupling tends to vanish with the distance between impurities much faster than $1/r^3$ expected for graphene monolayer. This can be observed both for different sublattices (Fig.~\ref{fig:fig3}(a)) as well as for the same sublattice (Fig.~\ref{fig:fig3}(b)). The dependence shows signs of exponential decay, according to $J^{RKKY}\propto \exp\left(-r/R\right)$. This kind of behaviour is observed for every distance of the impurity pair from the AGNR center, i.e. for every $\delta$; only some magnitude variations are present. Such an exponential decay is characteristic of the systems which indicate a gap in the electronic spectrum. For example, it has been mentioned in the Ref.~\onlinecite{Dugaev2006} that the gap opening in graphene due to spin-orbit interactions should lead to such a behaviour. This kind of coupling is also known from the semiconductor physics as Bloembergen-Rowland coupling \cite{Bloembergen1955} mediated by the excited electron-hole pairs. The characteristic decay distance $R$ is energy gap-dependent and faster decay is predicted for narrower AGNRs with the larger energy gap. For the next width, $N=9$ the AGNR is also semiconducting. Despite the small difference in the energy gap value, the distance behaviour of the coupling is noticeably different. Namely, the decay with the distance appears significantly slower.

The situation is very different for the metallic AGNR (studied case of $N=11$). From the plots Fig.~\ref{fig:fig3}(e),(f), it can be concluded that the coupling behaves in a manner characteristic of semiconducting systems only for the impurities distance from the AGNR symmetry axis $\delta$ such that $\delta \mod 3 = 0$. For the remaining distances $\delta$, including especially the impurities at the AGNR edge, the distance dependence is much weaker. As it can be observed, for different sublattices, the AF interaction tends to vanish quite slowly, and it can be numerically checked that it follows the $J^{RKKY}\propto 1/r$ rule. Such a dependence is rather characteristic of metallic-like, one-dimensional systems.\cite{Kittel1968,Litvinov1998,Aristov1997,Giuliani2005}. The distance dependence is even more interesting for the impurities at the same sublattice. Let us observe that the interaction energy, after some initial decrease, is almost distance-independent. Moreover, the coupling magnitude for this case is greatly pronounced in comparison with the other cases. 

This peculiar behaviour of the indirect coupling can be understood owing to the analysis of the zero-energy states wavefunctions for AGNRs. Let us consider first the case of two magnetic impurities in different sublattices, situated such that $\delta \mod 3 \neq 0$. In such a situation, each one of the two zero-energy states is occupied by a single electron, with the spin orientation which minimizes the value of the term $S_{k}s^{z}_{k}$ present in the Hamiltonian Eq.~(\ref{eq:eq1}). The first-order perturbation calculus correction to the energy of each state comes from the interaction with one impurity only and is equal to $\Delta E_{0}=-\frac{1}{2}n_{0}S\left|J\right|$, thus does not depend on the relative orientation of both impurity spins. Therefore, in such a situation zero-energy states give no first-order contribution to the indirect coupling. Let us also note that the zero-energy state is two-fold degenerate (not including spin degeneracy). The matrix elements of the Anderson-Kondo term taken between two different zero-energy states vanish. As a consequence, the basis of the zero-energy states associated each with different sublattice constitutes a proper basis for first-order perturbation calculus including the degeneracy. The second-order contribution emerges due to non-vanishing matrix elements of the Anderson-Kondo term taken between one of the zero energy states and other states.

The situation is quite different when both magnetic impurities are situated in the same sublattice. If the impurity spins are aligned ferromagnetically, the zero-energy state belonging to this sublattice is occupied by a single electron with the spin direction which minimizes its energy. The remaining state (belonging to the other sublattice) is also occupied by a single electron, but it is unperturbed and its energy is unchanged. As a consequence, the first-order correction to the energy is $\Delta E^{F}_{0}=-n_{0}S\left|J\right|$. On the other hand, it the impurity spins are polarized antiferromagnetically, the first-order corrections to the energy of the state associated with their sublattice cancel each other. The other zero-energy state is still unperturbed, and therefore $\Delta E^{AF}_{0}=0$. Taking in consideration the formula Eq.~(\ref{eq:eq2}), we obtain the indirect exchange integral associated with zero-energy states (in first-order perturbation calculus) equal to $J^{RKKY}=\left(\Delta E^{AF}-\Delta E^{F}\right)/2S^2$, what leads to 

\begin{equation}
\label{eq:eq3}
J^{RKKY}=n_{0}\left|J\right|/2S.
\end{equation} 

Let us emphasize that this kind of contribution to the indirect coupling is absent when any state is filled with two opposite-spin electrons, since then the first order correction to the energy of the state cancel out, while the second order corrections give rise to an ordinary RKKY indirect coupling mechanism, with interaction proportional to $J^2$. The first-order mechanism bears some resemblance to double exchange and is somehow analogous to the ferromagnetic contribution to indirect coupling found in ultrasmall graphene nanoflakes \cite{Szalowski2011}. Due to the proportionality to $\left|J\right|$, it strongly dominates over the typical second-order mechanism for weaker contact potentials $J$. It is worth emphasizing that, despite the linear dependence on the contact potential, the coupling is insensitive to the sign of $J$.

\begin{figure}
\includegraphics[scale=0.25]{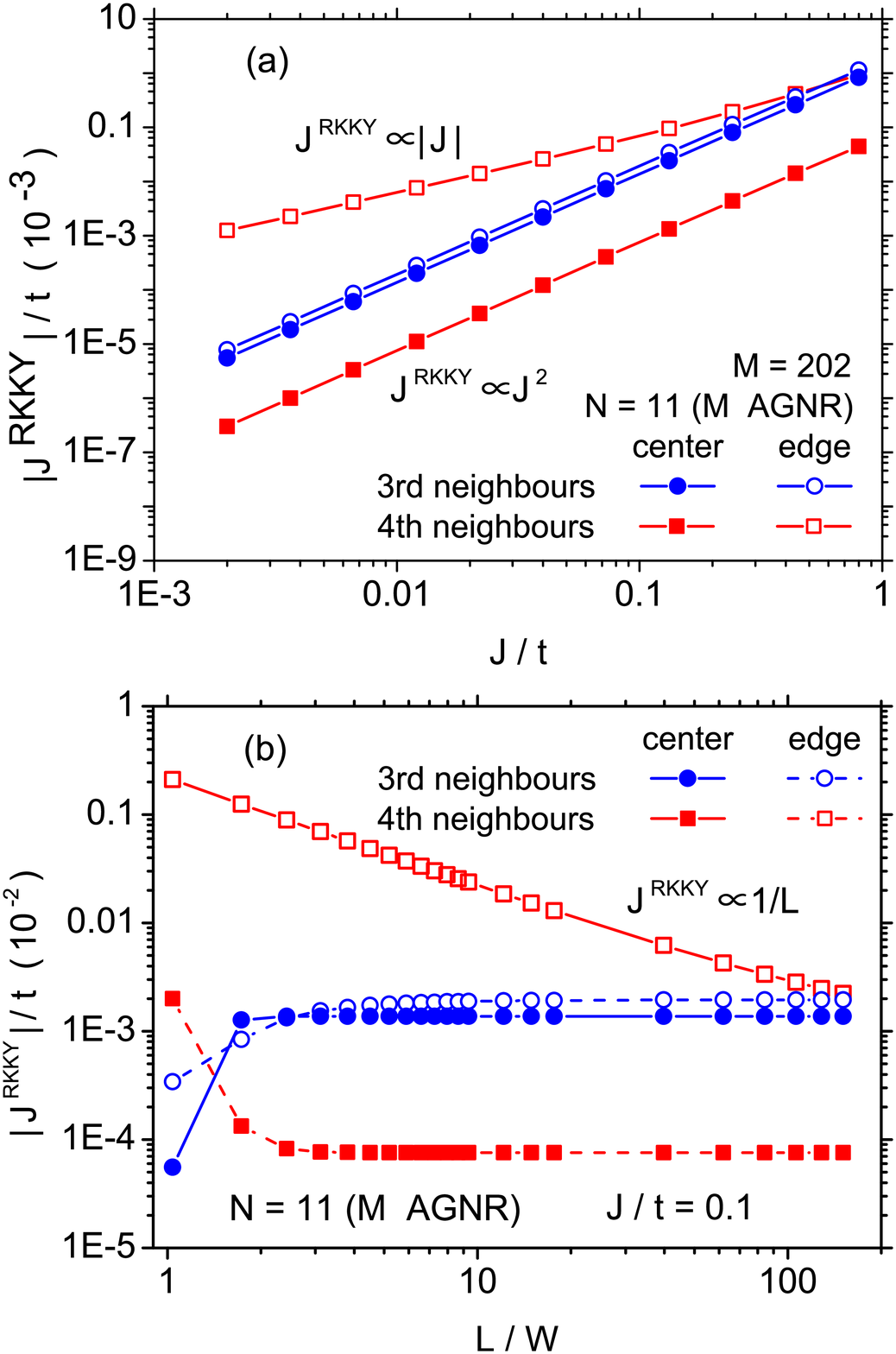}
\caption{\label{fig:fig4}(a) The dependence of the indirect coupling magnitude on the value of contact potential $J$, for the impurities being 3rd and 4th neighbours, placed in the center or at the edge of a M AGNR with $N=11$. (b) The dependence of the indirect coupling magnitude on the length-to-width ratio for M AGNR with $N=11$, for the impurities being 3rd and 4th neighbours, placed in the center or at the edge.}
\end{figure}

\begin{figure*}
\includegraphics[scale=0.25]{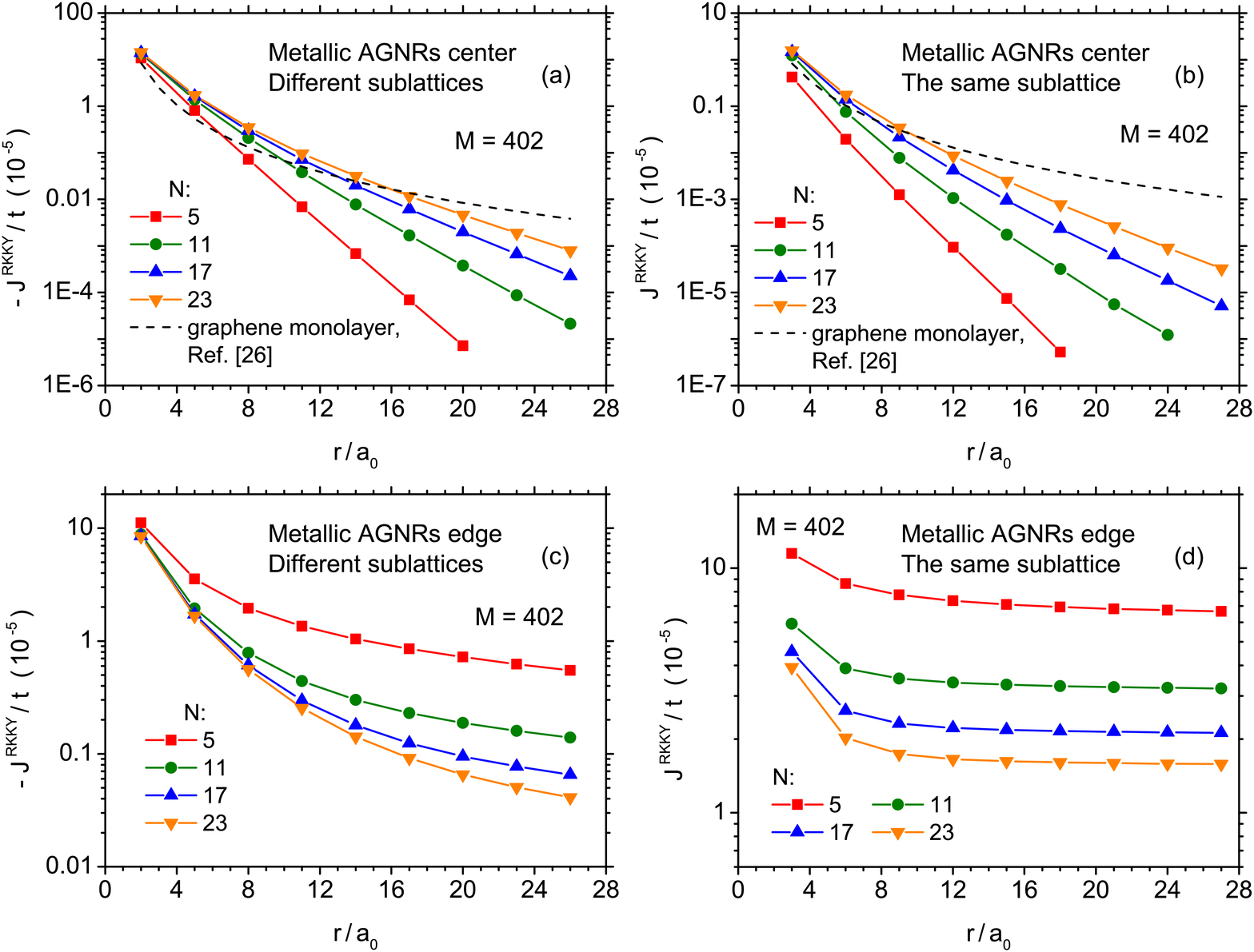}
\caption{\label{fig:fig5}Distance dependence of indirect coupling magnitude for 4 selected metallic AGNRs (of widths from $N=5$ to $N=23$), for impurities in different sublattices (left) and in the same sublattice (right), for impurities placed in the center and at the edge of the nanoribbon. Dashed lines denote the results of perturbative calculation after the Ref.~\onlinecite{Sherafati2011}. The plots are presented for various distances of the impurities from the AGNR center, covering the whole range between the center and the edge of the nanoribbon.}
\end{figure*}

In order to illustrate this, let us study the dependence of coupling magnitudes on the interaction energy $J$ between the localized spins and charge carrier spins. The results of the calculations are depicted in double logarithmic scale, in the Fig.~\ref{fig:fig4}(a), for a selected metallic AGNR with $N=11$. The magnetic impurities were placed either at the edge or on the symmetry axis and belonged to the same or different sublattices. For the impurities in different sublattices, it is visible that the coupling energy $J^{RKKY}$ is proportional to $J^2$. This means that the indirect interaction can be described using the second-order perturbation calculus (which is typically used to study indirect, charge-carrier mediated coupling in various systems). The situation is different when the impurities belong to the same sublattice. Then, the quadratic dependence of $J^{RKKY}$ on $J$ is conserved for the impurity pair in the AGNR center. In contrast, the interaction between edge impurities shows linear dependence of magnitude on $|J|$. This indicates that such an interaction contains a dominant contribution describable by means of first-order perturbation calculus, as it has been pointed out above with reference to the zero-energy states. The interaction magnitudes agree with those calculated from the Eq.~(\ref{eq:eq3}). In the studied range of $J$ values, the coupling between the edge impurities is at least an order of magnitude stronger than for center impurities.

It is of particular importance that the first-order contribution to the coupling is inversely proportional to the number of carbon atoms in the AGNR, thus it vanishes for infinite length of the nanoribbon and essentially constitutes a finite-size effect. This is a result of existence of just two zero-energy states for the whole system, for which $n_0=3/\left(M\cdot N\right)$.  

The importance of the finite length of AGNR for the mentioned ferromagnetic contribution to the coupling can be studied on the basis of Fig.~\ref{fig:fig4}(b). This figure presents the indirect coupling magnitude as a function of $L/W$ ratio for metallic AGNRs having the width of $N=11$, for impurities in the same or in different sublattices, in AGNR center or at the edge. It is clear that for centrally placed impurities and as well for edge impurities in different sublattices, the coupling energy is independent on AGNR length for $L/W$ ratios $\gtrsim 10$ (which means achieving the limit of an infinite system size). In contrast, for edge impurities in the same sublattice, the dependence of the type $J^{RKKY}\propto 1/L$ is clearly observed. Note that even for $L/W$ over 100, the coupling for edge impurities exceeds the one for center impurities by approximately 2 orders of magnitude. Such a slow decrease establishes the importance of this contribution to indirect RKKY coupling.

The evolution of distance dependence of indirect coupling when the metallic AGNR width is increased can be studied in Fig.~\ref{fig:fig5}. For impurities located in the center of the AGNR (Fig.~\ref{fig:fig5}(a),(b)), the coupling decay is faster than $1/r^3$ for all the illustrated AGNR widths, but the slope decreases gradually tending to reproduce the behaviour characteristic of an infinite graphene monolayer. This kind of evolution can be expected for semiconducting systems with a gap decreasing with the increase of the ribbon width. Recall that for the AGNR center, the zero-energy states are not involved in an indirect coupling. Totally different situation is met for the impurities at the AGNR edge, for which the position zero-energy states can contribute. For edge impurities in different sublattices (Fig.~\ref{fig:fig5}(c)), the $1/r$ dependence is characterisatic of the thinnest AGNR case (what, as mentioned, resembled RKKY coupling behaviour for one-dimensional metallic systems). It can be verified by plotting the dependence in double logarithmic scale (not shown) that the power decay law of coupling takes place for all studied widths of AGNR, however, the exponent varies from -1 toward -3 when $N$ increases. Therefore, the distance behaviour of RKKY coupling evolves toward the expected behaviour for infinite graphene monolayer, for which $J^{RKKY}\propto 1/r^3$ is predicted. Moreover, the magnitude of indirect coupling is significantly reduced as a result of increase of the width. On the other hand, when the edge impurities are located in the same sublattice (the case illustrated in Fig.~\ref{fig:fig5}(d)), the influence of the AGNR width on the coupling magnitude is most pronounced. Under such conditions a first-order contribution to the coupling from a zero-energy state is present, but its value decreases inversely proportionally to the AGNR width (i.e. to the number of the lattice sites in the AGNR whilst the length is kept constant). For all the studied AGNR widths, the distance dependence of the coupling is very weak for larger distances.

\begin{figure}
\includegraphics[scale=0.25]{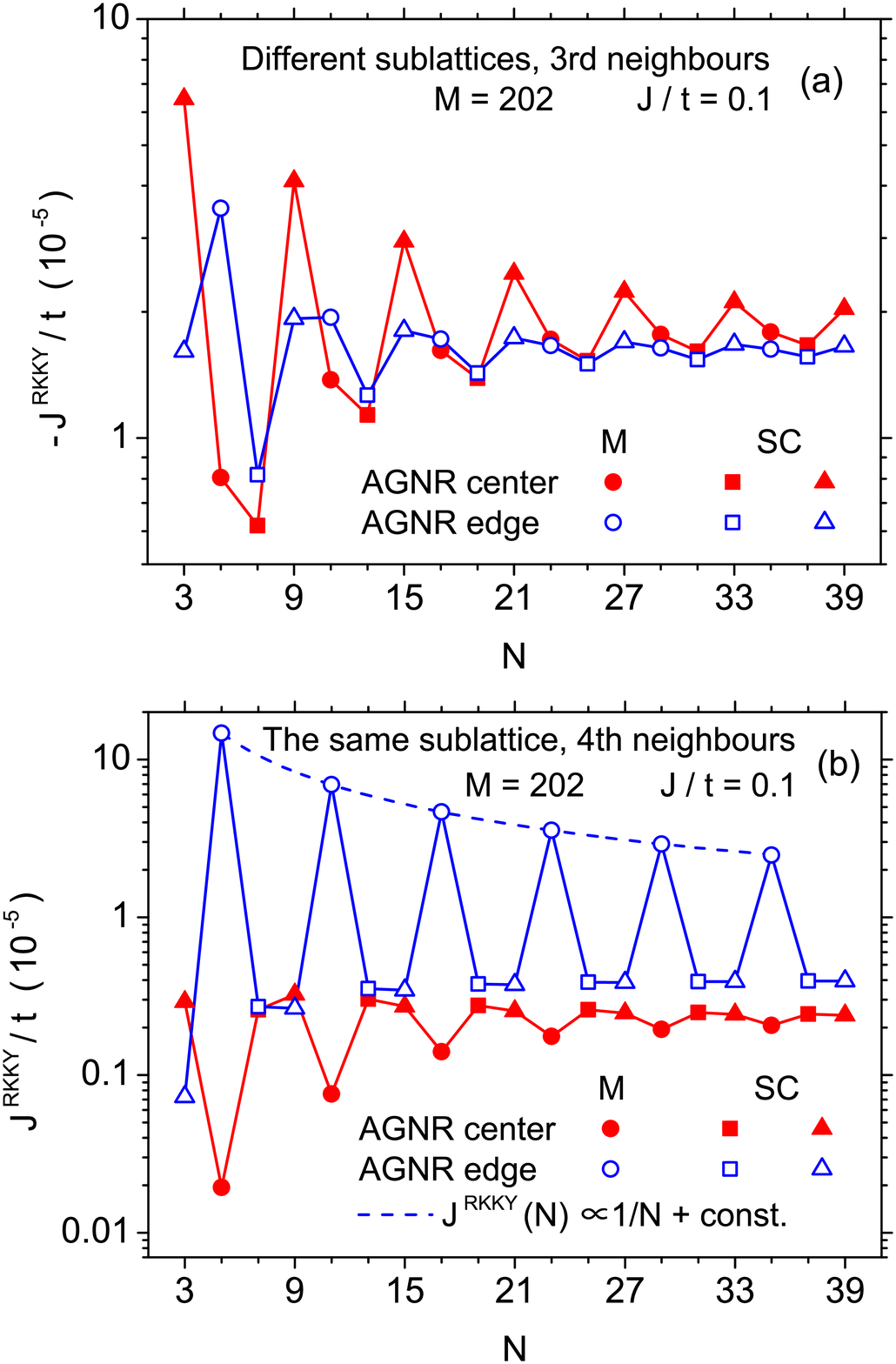}
\caption{\label{fig:fig6}The coupling between (a) 3rd neighbour impurities (b) 4th neighbour impurities as a function of the AGNR width $N$. The coupling magnitudes are presented for the center and edge of the AGNR. Values for metallic AGNRs and two kinds of semiconducting AGNRs are depicted with different symbols.}
\end{figure}

In order to analyse the width dependence on the coupling for the fixed distance between the impurities, let us focus on the case of the 3rd neighbours (situated in different sublattices) and the 4th naighbours (the same sublattice). These cases were illustrated in the Fig.~\ref{fig:fig6}, which presents the dependence of the coupling energy $J^{RKKY}$ for $J/t=0.1$ on the nanoribbon width (proportional to the number of atoms at the zigzag edge $N$). Fig.~\ref{fig:fig6} shows the coupling between the impurities being the 3rd neighbours, situated either on the symmetry axis or at the AGNR edge. It is visible that a pronounced dependence of the coupling energy on the AGNR width occurs, with an oscillatory envelope. The oscillations tend to vanish for wider AGNRs. A period equal to 3 is visible in the changes vs. $N$, what corresponds with the width dependence of the electronic structure of AGNRs, as studied using the tight-binding model by Wakabayashi\cite{Wakabayashi1999,Wakabayashi2009,Wakabayashi2010}. The pronounced quantum size effects in RKKY coupling magnitude follow this rule. It can be also deduced from the calculations for further neighbours that the oscillations envelope vanishes slower with the width for more distant impurities. Fig.~\ref{fig:fig6}(b) depicts the case of the impurities being the 4th neighbours, at the edge or in the center of the AGNRs. The quantum size effects for the impurity pair in the center resemble much ones observed for the previous case, with the same period of 3. The situation is much different for the edge impurities. There, a strong enhancement of the ferromagnetic coupling is observed each time when $N$ corresponds to the metallic AGNR. The origin of this coupling has been previously discussed. Let us observe that the energy of this coupling for metallic AGNRs varies according to $J^{RKKY}\propto 1/N+\mathrm{const.}$, in agreement with the described mechanism. The quantum size effects for AGNRs were already studied for example for the excitonic spectra \cite{Alfonsi2012,Jia2011,Mohammadzadeh2011}.

\begin{figure}
\includegraphics[scale=0.25]{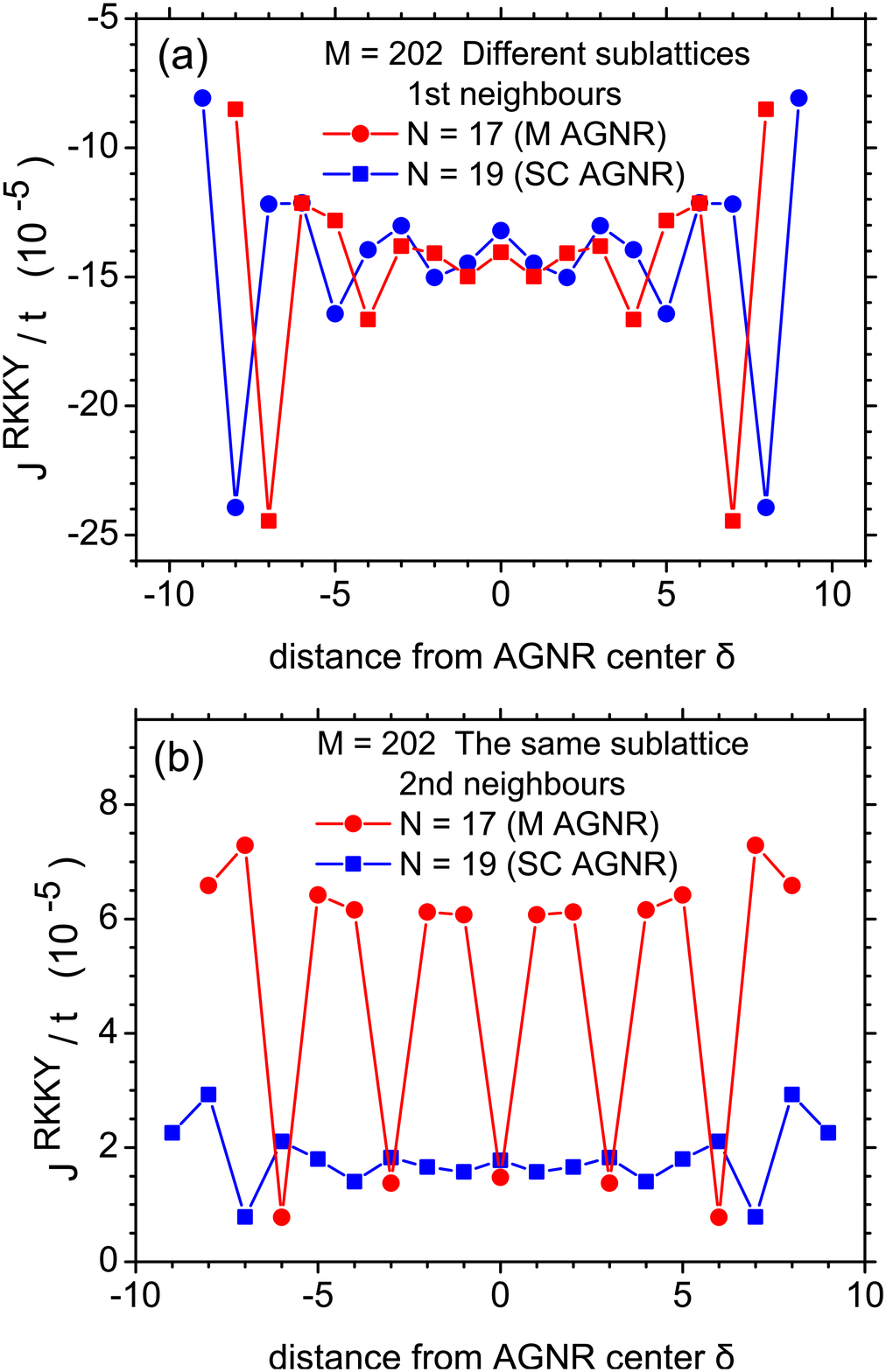}
\caption{\label{fig:fig7}The dependence of the indirect coupling on the distance of the impurity pair from the AGNR center, plotted for the impurities being 1st neighbours (a) and 2nd neighbours (b), for two selected AGNRs - a SC one and a M one.}
\end{figure}

Finally, let us concentrate on the variations of the coupling magnitude for selected impurity pairs for various pair distances $\delta$ from the AGNR symmetry axis. Such calculation results are presented in Fig.~\ref{fig:fig7}. For the 1st neighbours [Fig.~\ref{fig:fig7}(a)], the coupling behaves similarly for metallic and semiconducting AGNR, indicating pronounced oscillations while moving away from the symmetry axis towards the edge. In the vicinity of the edge the variations gain particularly large magnitude. This distribution of the RKKY coupling energies across the width of an AGNR is similar to Friedel oscillations, which are ubiquitous in numerous physical properties of the systems with boundaries (e.g. \cite{Pelc2008a,Pelc2008b}). For the 2nd neighbours [Fig.~\ref{fig:fig7}(b)], the difference between metallic and semiconducting AGNR is clearly emphasized. For SC AGNR, the dependence of coupling magnitude resembles the Friedel oscillation-like picture from Fig.~\ref{fig:fig7}(a). On the contrary, due to the existence of the mechanism enhancing the ferromagnetic coupling with the help of the zero-energy state in M AGNR, the interaction is strongly pronounced for $\delta\neq 3d$ (as explained before). Thus, the coupling energy is greatly pronounced except at that atomic rows for which the zero energy state charge density vanishes. Let us note here that this enhancement would decrease in magnitude for longer AGNRs.

\section{Final remarks}

In the paper, the indirect (RKKY-like) coupling between on-site magnetic impurities localized in an armchair graphene nanoribbon has been studied. The non-perturbational method, based on the total energy calculation, allowed for capturing not only the usual RKKY mechanism attributed to second-order perturbation calculus, but also a first-order mechanism. It is worth emphasizing that the magnitude of the indirect RKKY coupling between magnetic impurity spins in AGNRs is strongly dependent on the location of the impurities in the nanoribbon, and the coupling energy exhibits pronounced oscillatory behaviour resembling the Friedel oscillations. When the indirect coupling is of typical origin, its distance dependence is predominantly of exponential kind, similar to Bloembergen-Rowland coupling in gapped systems. This kind of behaviour appears also for metallic AGNRs when zero-energy states do not give contribution. 

On the contrary, the first-order mechanism becomes operative for the electronic states occupied only by a single electron, which can be formed in metallic nanoribbons. Due to a peculiar form of the wavefunction for this states, the leading contribution to indirect coupling can be either slowly decaying (one-dimensional metallic-like in character) for different sublattices, or constant for the same sublattice. The mentioned effects are inevitably connected with a finite size of the system since their presence is owing to the two zero-energy states only. However, even for considerably large length to width ratio for AGNR, they could give a noticeable contribution to the coupling. It could be worth noticing that the involvement of zero energy states in charge carriers mediated coupling implies the large sensitivity of the interaction to the number of charge carriers in the system. This feature in principle might open the door to control over the coupling sign and magnitude in AGNRs and may stimulate further studies of such systems. On the other hand, also the influence of external fields on the system might be of interest. 

Finally, let us note that the presence of zero-energy degenerate states is characteristic of a wider class of carbon nanostructures (e.g. \cite{Pelc2011}) and thus such effects may not be limited to the system studied here.

\begin{acknowledgments}
The computational support on Hugo cluster at Department of Theoretical Physics and Astrophysics, P. J. \v{S}af\'{a}rik University in Ko\v{s}ice is gratefully acknowledged.

This work has been supported by Polish Ministry of Science and Higher Education by a special purpose grant to fund the research and development activities and tasks associated with them, serving the development of young
scientists and doctoral students.

The author is deeply grateful to T. Balcerzak for critical reading of the manuscript.

\end{acknowledgments}

\end{document}